# Fully self-referenced frequency comb consuming 5 watts of electrical power

**Paritosh Manurkar,[1] Edgar F. Perez,[1] Daniel D. Hickstein,[2] David R. Carlson,[2] Jeff Chiles,[1] Daron A. Westly,[4] Esther Baumann,[1] Scott A. Diddams,[1] Nathan R. Newbury,[1] Kartik Srinivasan,[4] Scott B. Papp,[2,3] and Ian Coddington[1,*]**

[1]*Applied Physics Division, National Institute of Standards and Technology, 325 Broadway, Boulder, Colorado 80305, USA*
[2]*Time and Frequency Division, National Institute of Standards and Technology, 325 Broadway, Boulder, Colorado 80305, USA*
[3]*Department of Physics, University of Colorado, 2000 Colorado Avenue, Boulder, Colorado 80309, USA*
[4]*Center for Nanoscale Science and Technology, National Institute of Standards and Technology, 100 Bureau Drive, Gaithersburg, Maryland 20899, USA*
*\*ian.conddington@nist.gov*

**Abstract:** We present a hybrid fiber/waveguide design for a 100-MHz frequency comb that is fully self-referenced and temperature controlled with less than 5 W of electrical power. Self-referencing is achieved by supercontinuum generation in a silicon nitride waveguide, which requires much lower pulse energies (~200 pJ) than with highly nonlinear fiber. These low-energy pulses are achieved with an erbium fiber oscillator/amplifier pumped by two 250-mW passively-cooled pump diodes that consume less than 5 W of electrical power. The temperature tuning of the oscillator, necessary to stabilize the repetition rate in the presence of environmental temperature changes, is achieved by resistive heating of a section of gold-palladium-coated fiber within the laser cavity. By heating only the small thermal mass of the fiber, the repetition rate is tuned over 4.2 kHz (corresponding to an effective temperature change of 4.2 °C) with a fast time constant of 0.5 s, at a low power consumption of 0.077 W/°C, compared to 2.5 W/°C in the conventional 200-MHz comb design.

## 1. Introduction

There exists an increasing need to operate femtosecond frequency combs outside the laboratory [1,2]. Applications such as comb spectroscopy [3,4], femtosecond timing dissemination [5], low-noise microwave generation [6,7], and portable optical clocks [8,9] all require practical, fieldable frequency combs. While chip-scale combs may one day be a practical alternative [10], the current solution for fieldable combs remains fiber frequency combs with polarization-maintaining (PM) erbium (Er) fiber [11–14]. Self-referenced fiber frequency combs have already been demonstrated in moving vehicles [15], sounding rockets [16], industrial environments [17] and remote field sites [18]. As these combs become more robust and immune to environment perturbations, the principle challenge of fielding a comb shifts to the total electrical power consumption and the heat dissipation associated with their operation. Such issues might go unnoticed in the laboratory, but they constitute serious limits in the field. A typical fully self-referenced frequency comb will require two to four high-powered pump diodes for the erbium-doped fiber amplifier (EDFA) to produce the high-energy pulses traditionally needed for self-broadening. Additionally, careful temperature control of the oscillator is required to stabilize the comb repetition rate ($f_{\text{rep}}$) otherwise the thermal refractive coefficient of fiber (~$10^{-5}$/°C) would overrun all other cavity length modulations. Overall, it is not uncommon for a typical Er fiber comb to draw > 40 W of electrical power. This power draw is particularly cumbersome in space, where electrical power is both expensive and limited [16,19]. As a point of reference, a CubeSat may have only 30 W of electrical power to distribute between all systems.

For the comb design demonstrated in Ref. [15], which is typical of many fiber systems, the power draw is equally divided between temperature control of $f_{rep}$ and generation of the carrier-envelope offset ($f_{CEO}$) signal for self-referencing. For $f_{CEO}$ generation, the aforementioned 200-MHz comb requires almost 2 W of pump power corresponding to almost 20 W of conditioned, DC electrical power. Temperature control is often applied to a large enclosure when only the oscillator fiber at the heart of the comb needs to be controlled. Here, we demonstrate two solutions to reduce the power consumption of a frequency comb by almost a factor of 10. First, by performing spectral broadening in silicon nitride ($Si_3N_4$, henceforth SiN) waveguides [20–22], we can greatly reduce the optical power required for self-broadening by a factor of 5. At the same time, we have redesigned the waveguide to allow efficient spectral broadening with 200-fs pulses as opposed to <80-fs pulses used in Ref. [22]. There are two important consequences of this redesign. First, <80-fs pulses from a fiber system are typically generated via nonlinear amplification in a high-gain, normal-dispersion fiber amplifier [23]. (For instance, in Ref. [22], pulses are amplified to >2 nJ before being attenuated to ~100 pJ). Broadening with 200-fs pulses greatly increases the design space available for the combs, allowing for low power linear amplifiers and truly low power operation. Secondly, the use of low gain fiber amplifiers means one can practically use low power, passively-cooled pump diodes for these amplifiers. Since active cooling can consume up to two-thirds of power draw of a pump diode, it further magnifies the energy savings from the SiN approach. As an additional point of interest, low power fiber amplifiers can be constructed from lightly doped erbium fiber, which can be made more compatible with a radiation environment in space.

The second principle source of energy consumption in a frequency comb is temperature control of the oscillator, which is necessary for any frequency stabilized comb. This temperature control is typically achieved by controlling the oscillator enclosure with heaters or thermoelectric coolers, often consuming several Watts of power. Typically, one will limit this power draw by compressing the size of the oscillator package to reduce the thermal mass to be controlled. Here, we take this idea to the logical extreme and use metalized fiber — as first employed in the telecom industry [24–26] — to resistively heat and control only the very small thermal mass of the optical fiber in the oscillator. By direct resistive heating of intra-cavity gold-palladium-coated fiber, which we refer to as a "fiber resistive modulator," we demonstrate a simple and low-power temperature tuning solution consuming ~77 mW/°C.

Together these two techniques allow for a 5–10× reduction in power consumption and bring the total consumption of the comb system below 5 W. At this power, a host of new applications should be possible including battery powered operation.

## 2. *f*-to-*2f* self-referencing at low power

The low power comb design developed here is shown in Fig. 1(a). The oscillator is based on the Er-doped fiber design, presented in Ref. [11], but is modified to operate at an $f_{rep}$ of 100 MHz. The cavity consists of all anomalous dispersion, PM fiber with semiconductor saturable absorbing mirror (SESAM) based modelocking. Output coupling is generated with a dielectric coated FC/PC connector with 15% transmission. The oscillator outputs a modest 11-nm spectral bandwidth centered at 1562.6 nm, 200-fs pulse train, and an average power of 2 mW. Amplification to 34 mW is performed in a 45-cm segment of Er doped fiber, which provides ~12 dB gain and roughly maintains the pulse characteristics (bandwidth of 12.5 nm, and 170-fs pulses at the output). This modest output power and pulse width are achieved by pumping the oscillator and amplifier with efficient, passively-cooled 250-mW pump diodes at 980 nm. Each of these diodes is designed to consume 2.5 W of electrical power.

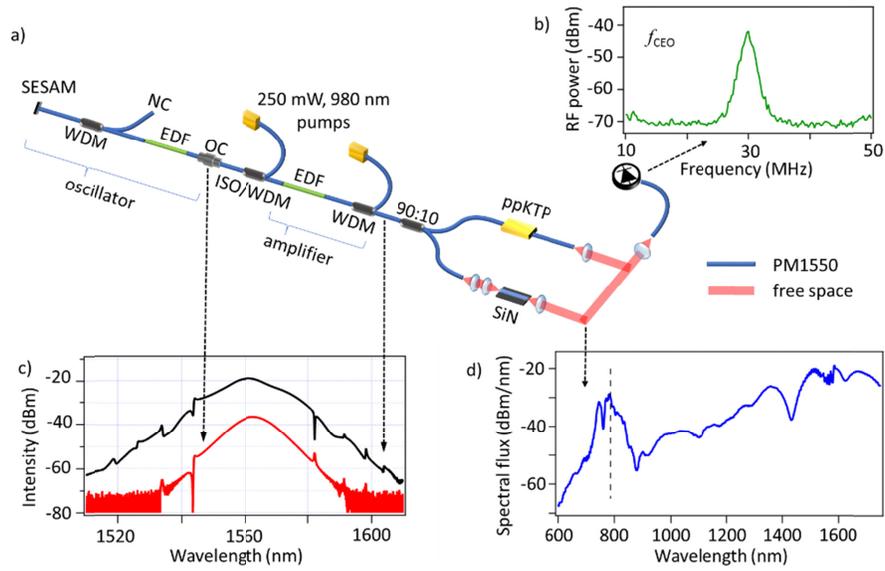

Fig. 1. (a) Layout of the low power comb. EDF: erbium-doped fiber, OC: output coupler, WDM: wavelength division multiplexer, ISO: isolator, PPKTP: periodically-poled potassium titanyl phosphate, SiN: silicon nitride, NC: not connected. Note that this oscillator is pumped through the end mirror (which is transparent at 980 nm) using the passively-cooled laser diode at 980 nm. The intracavity WDM only serves to dump extra pump light, protecting the SESAM and limiting back reflection into the pump. After the amplifier, a coupler directs 90% of the output to the SiN waveguide for broadening and 10% to a PPKTP crystal for doubling. (b) 30-dB $f_{CEO}$ signal recorded with 22 mW of light incident on the SiN waveguide (300-kHz resolution bandwidth). (c) Spectra of the comb at the output of the oscillator (red), at the output of the amplifier (black). (d) Spectra at the output of the SiN waveguide to beat against the doubled light from PPKTP.

Spectral broadening is achieved via supercontinuum generation in an SiN waveguide fabricated by Ligentec [27] using the "Photonic Damascene" process [28,29]. The waveguide thickness (height) is 750 nm, the width is 2100 nm, and the total length is 21.05 mm. The waveguide is fully clad with silicon dioxide ($SiO_2$). Inverse tapers on the input and output facets enlarge the mode and provide better coupling, which is estimated at −2 dB per facet. The waveguide provides anomalous dispersion at the 1560-nm oscillator wavelength, which leads to supercontinuum generation via the soliton fission process [30]. Such waveguides have been shown to provide octave-spanning spectra with as little as 100-pJ pulses incident on the waveguide facet [20–22]. Here, we require ~200-pJ pulses due to the somewhat longer pulse duration. However, it is likely that a longer and dispersion optimized waveguide would allow for broadening with as little as 100 pJ given our ~200-fs pulse width.

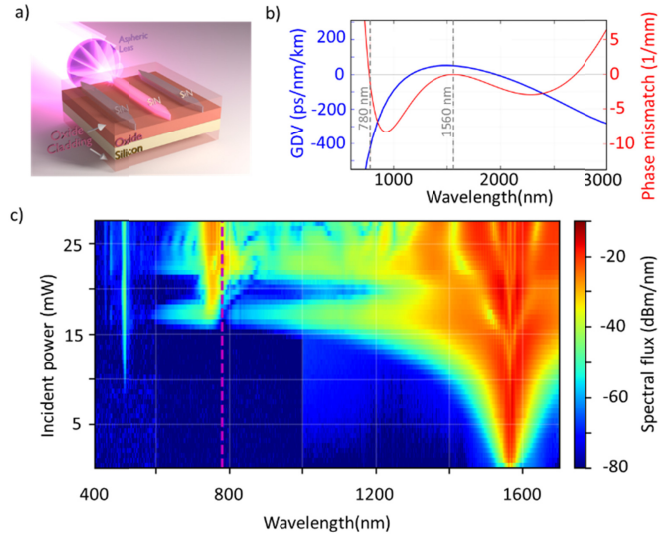

Fig. 2. (a) SiN chip used in this experiment. (b) Calculated dispersion profile for the 2100-nm waveguide. (c) recorded spectrum vs incident power. The dotted line denotes 786 nm.

With 23 mW of comb light incident on the waveguide, we output roughly 156 μW of power within the "dispersive wave" centered at ~760 nm (>1% of the total output) while depositing relatively little optical power in other, unused, spectral bands, as seen in Fig. 1(d). The dispersive wave is stable for a large range of input powers and its center wavelength can be tuned by optimizing physical parameters of the SiN waveguide. The $f$-to-$2f$ signal is generated by doubling 1572-nm light from the amplifier in a periodically-poled potassium titanyl phosphate (PPKTP) waveguide, combining it with the supercontinuum light in a single mode fiber and detecting on a silicon avalanche photodiode. When pumped with 2.5-mW pulses, centered at 1560 nm, the PPKTP generates 12 μW of light in a 6-nm band which, when mixed with ≈2.7 μW of supercontinuum light in the same band yields an $f_{CEO}$ signal with a 30-dB signal-to-noise ratio in a 300-kHz bandwidth [Fig. 1(b)]. It is worth noting that while the powers used here are low, there is still further room for optimization. As can be seen in Fig. 2(c), the SiN waveguide spectra is peaked at wavelengths shorter than 786 nm, a future waveguide with slightly higher zero dispersion crossing or a PPKTP waveguide phase matched for 760 nm would further improve the system efficiency.

In our current configuration, the entire power draw of the pump diodes—including diode drivers — is 4.6 W, well below the power draw of common battery-powered devices (an operating laptop will consume between 20 and 100 W). Indeed, we successfully powered the comb off a handheld USB charger for more than an hour while fully self-referenced. Figure 3 shows the counted $f_{rep}$, and $f_{CEO}$ during this duration. (The photodetectors and locking electronics were powered separately).

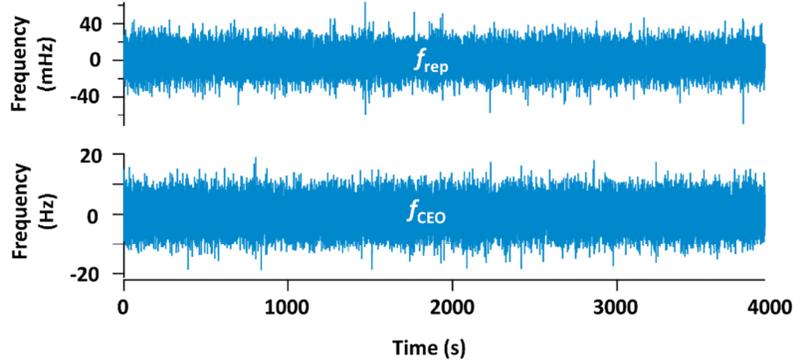

Fig. 3. Counted $f_{rep}$, and $f_{CEO}$ while the frequency comb was locked and powered by a USB charger. Counter gate time was 100 ms. The standard deviations calculated for the first 1000 points are 12.7 mHz for $f_{rep}$ and 4.7 Hz for $f_{CEO}$.

Space applications have an obvious need for low power combs. Our design provides that low-power alternative and presents some additional advantages from a radiation hardness perspective. While standard PM fibers are available in a radiation-hardened variety, radiation-hardened highly-doped Er PM fibers are more difficult to obtain. Here, the lower peak-power required for self-broadening relaxes the gain and dispersion requirements for the Er-doped fiber and makes this design more compatible with the available radiation-hardened fibers. Secondly, replacing traditional highly-nonlinear fiber with the SiN waveguide is attractive since the properties of nonlinear fiber when subjected to radiation are unknown whereas SiN has been exploited as a radiation hardened material for decades in electronics [31,32] and recent work has shown that this robustness extends to its optical properties as well [33].

## 3. Fiber resistive modulator

To demonstrate low-power temperature stabilization and tuning of the oscillator, we incorporated a 13.9-cm-long fiber coated with an alloy of Gold-Palladium (AuPd, 47% Au and 53% Pd) into the cavity. Electrical current traveling through this coating modulated the cavity length (hence $f_{rep}$) by direct resistive heating of the fiber. To apply the AuPd coating, the buffer coating on the fiber was stripped off so that the metal layer adhered directly to the 125-μm diameter fiber cladding. Here, we applied the coating to the Er-doped fiber portion of the cavity but the AuPd coating could equally well be applied to the intra-cavity PM1550 fiber of any length. For coating, the fiber was wound in a spiral pattern and mounted on the surface of a silicon wafer, such that the fiber did not cross over itself at any point. After mounting, the fiber was cleaned with oxygen plasma ashing to remove any organic residues on the surface. We then utilized electron-beam evaporation to deposit 10 nm of titanium (Ti) followed by 100 nm of the AuPd alloy on the exposed surface of the fiber, effectively coating roughly half of its surface area. The initial Ti layer improved the adhesion of the AuPd layer. We selected AuPd instead of Au because of the former's higher resistivity, which ensures heating of the fiber with less ohmic loss in the connecting leads. The total resistance of the 13.9-cm coated fiber was 2.9 kΩ.

To demonstrate the fiber resistive modulation of $f_{rep}$, we made an electrical connection to each end of the AuPd coating with an electrically-conducting epoxy. A voltage of 1 V was then applied across the connections and increased in steps of 1 V while monitoring the current draw and $f_{rep}$. Figure 4 shows the measured $f_{rep}$ as a function of the electrical input power to the AuPd coating. At the 30-V limit of our supply, 322 mW of power was consumed by the resistive heater while yielding a 4.2-kHz shift in $f_{rep}$. Higher modulation depths would certainly be possible with higher voltage or a lower resistance modulator. Since the thermal refractive coefficient of fiber is ~$10^{-5}$/°C, the 4.2-kHz shift corresponds to an average oscillator temperature change of $\Delta T \approx 4.2$ °C. Of course, the resistive heating takes place only in a 13.9-cm coated section of the cavity, which has a much higher temperature change of 4.2°C × 101.9 cm/13.9 cm = 30.8 °C (where cavity length = 101.9 cm).

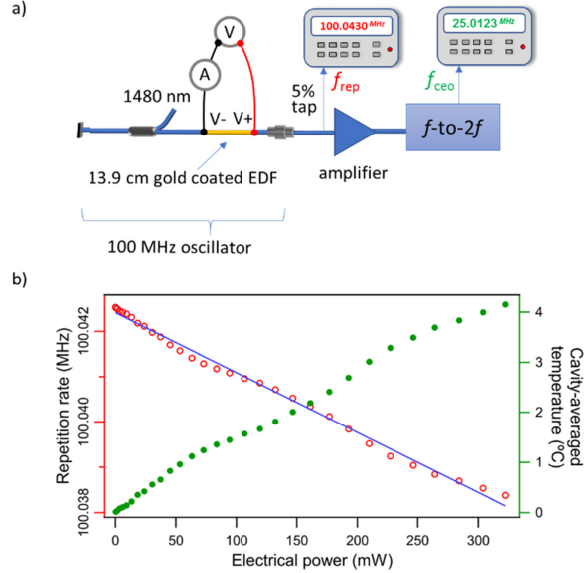

Fig. 4. Fiber resistive modulator design and results. (a) Oscillator design: in this case, only 13.9 cm of the ≈1 m oscillator is coated. (b) Repetition rate (left axis) and cavity-averaged temperature (right axis) versus input electrical power. Blue line is a linear fit giving an $f_{rep}$ tuning constant of 77 mW/kHz. EDF: Erbium-doped fiber, $f_{rep}$: repetition rate, $f_{CEO}$: carrier envelope offset frequency.

To measure the response time constants ($\tau$) of the fiber resistive modulator, we applied an electrical square-wave modulation at 6 V while monitoring the repetition rate. A fit to the rising and falling edges of the response curves with exponentials yields time-constants $\tau_{rise}$ and $\tau_{fall}$ of 0.57 s and 0.47 s, respectively, as shown in Fig. 5(a). The inset of Fig. 5(a) also shows the response of $f_{CEO}$. To find the complete response transfer function (Bode plots), we averaged 500 such traces (to reduce the noise), separated the averaged rising and falling edges, padded and apodized each, and then Fourier transformed the result to find the magnitude and phase of the transfer function, as shown in Fig. 5(b). These give a 45° phase shift at 0.4 Hz.

We can also calculate the fixed point for the fiber resistive modulator based on the method described in Ref. [34]. using the data in Fig. 5(a) and inset. The fixed point, $f_{fix}$, is calculated as

$$f_{fix} = \frac{df_{CEO}}{df_{rep}} f_{rep} - f_{CEO}, \qquad (1)$$

which yields 4.5 THz, similar to most fiber stretchers as would be expected.

The principle advantage of the fiber resistive modulator approach is that it allows for compensation of ambient temperature swings with modest power consumption, but there are also comb design advantages provided by the speed of this approach. Typically, cavity length modulation in a fiber oscillator is delegated to several modulators. Fast modulation, >1 Hz, is often handled by a fast small throw PZT or EOM [35,36]. Slow modulation to counteract temperature drift is handled by temperature feedback or translating optics typically having several seconds to minute response times. Often an additional, large dynamic range PZT modulator, is required to bridge the gap in timescales between fast and slow modulators.

Direct resistive heating of the fiber offers sufficient speed and dynamic range to potentially replace both slow and intermediate range modulators while requiring no free space alignment or bulky associated optics, thus reducing system complexity, size, and weight.

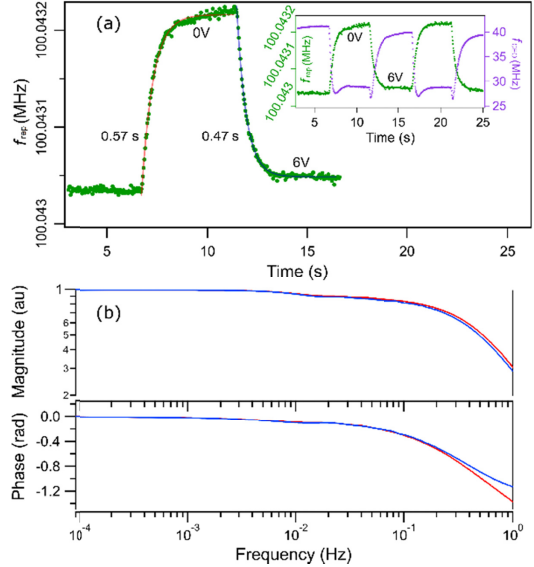

Fig. 5. Response time constants and Bode plots. (a) The response of $f_{rep}$ (green) and $f_{CEO}$ (shown in the inset in purple) to a square wave voltage modulation. The exponential fits (transparent red and violet lines) extracted the time constants ($\tau$) of the $f_{rep}$ response. (b) Transfer function (Bode plots) for rising (red) and falling (blue) edges.

## 4. Discussion

Here, we demonstrate a novel low-power frequency comb design that greatly reduces the power requirements of a fully self-referenced comb. Table 1 shows a comparison of a typical comb from Ref. [15] and two power optimized 100-MHz versions — one including only the fiber resistive modulator design and a second that is fully optimized with passively-cooled pumps for oscillator and amplifier, and SiN waveguides for self-broadening. The stated power consumptions include the entire comb system excluding the power consumption from locking electronics. The power consumption has been further reduced by use of an optimally-designed PPKTP doubling waveguide in the $f$-to-$2f$ self-referencing scheme. These devices have been designed for optimal quasi-phase matching at a specific temperature and have a weak, ~10 °C/nm, temperature-tuning response at these wavelengths. For environments with known temperature that drift less than 10 °C, active temperature control should be unnecessary.

Though not chip scale, fully self-referenced hybrid fiber/waveguide frequency combs can be made compact and low power. While direct electrically-pumped combs could one day improve on these power consumption levels, the size and power draw shown here is already quite low. At these levels, the size and power consumption of any larger system employing the comb will likely be limited by the surrounding electronics and not the comb itself.

Table 1. Comparison of electrical power consumption between a conventional 200-MHz comb design (left column), and two optimized 100-MHz designs, first only with fiber resistive modulator (middle) and second fully optimized using passively-cooled pumps (right). It should be noted that the fully optimized design assumes the use of a fiber resistive modulator for temperature tuning. All the values are based on actual measurements. The temperature tuning values are based on a 3 °C offset from ambient temperature. Measured power draw for the pump diodes includes the diode controllers (Wavelength Electronics LDTC1040 and LDTC2E for the oscillator and amplifier pump respectively and two LDTC1024's for the passively-cooled pumps [27]). The power consumption of the locking electronics including the PZTs has not been included here, although for the PZT the consumption is negligibly low (< 5 mW)

| Parameter | Conventional 200-MHz comb (HNLF + PPLN) | 100-MHz comb + fiber resistive modulator + HNLF + PPKTP | 100-MHz comb + passively-cooled pumps + SiN waveguide + PPKTP |
|---|---|---|---|
| Temperature tuning of $f_{rep}$ | 7.5 W | 0.23 W | 0.23 W |
| Oscillator pump | 4.4 W | 4.4 W | 1.85 W |
| Amplifier pumps | 20 W | 10 W | 2.75 W |
| Doubling waveguide TEC | ~1 W [11] | 0 W | 0 W |
| TOTAL | 33 W | 14.6 W | 4.8 W |

## Acknowledgments

We thank Ivan Ryger and John H. Lehman for discussions about deposition of nichrome and gold on fiber optics, and Isaac Khader for assistance in the measurements. Work of U.S. Government, not subject to copyright.

## Disclosures

The authors declare that there are no conflicts of interest related to this article.